\documentclass[superscriptaddress,amsmath,amssymb,aps,notitlepage,twocolumn,floatfix]{revtex4-1}

\usepackage{cmbright}
\DeclareFontShape{OT1}{cmss}{m}{it}{<->ssub*cmss/m/sl}{}
\renewcommand{\rmdefault}{cmss}
\renewcommand{\sfdefault}{cmss}
\DeclareFontFamily{OT1}{cmbr}{\hyphenchar\font45 }
\DeclareFontShape{OT1}{cmbr}{m}{n}{%
	<-9>cmbr8
	<9-10>cmbr9
	<10-17>cmbr10
	<17->cmbr17
}{}
\DeclareFontShape{OT1}{cmbr}{m}{sl}{%
	<-9>cmbrsl8
	<9-10>cmbrsl9
	<10-17>cmbrsl10
	<17->cmbrsl17
}{}
\DeclareFontShape{OT1}{cmbr}{m}{it}{%
	<->ssub*cmbr/m/sl
}{}
\DeclareFontShape{OT1}{cmbr}{b}{n}{%
	<->ssub*cmbr/bx/n
}{}
\DeclareFontShape{OT1}{cmbr}{bx}{n}{%
	<->cmbrbx10
}{}

\usepackage[noindentafter]{titlesec}
\usepackage{titlesec}
\usepackage{lipsum}
\titleformat{name=\section}
{\normalfont\large\bfseries\MakeUppercase}{\MakeUppercase{\thesection}}{0pt}{}
\titleformat{name=\subsection}
{\normalfont\bfseries}{\thesection}{0pt}{}
\titlespacing{\section}{0cm}{0.7cm}{0.01cm}
\titlespacing{\subsection}{0cm}{0.45cm}{0cm}

\usepackage[utf8]{inputenc}
\usepackage{listings}
\usepackage{footmisc}
\usepackage{enumerate}
\usepackage{latexsym}
\usepackage{braket}
\usepackage[version=4]{mhchem}
\usepackage{graphicx}
\usepackage[colorlinks=True,linkcolor=red,citecolor=blue,urlcolor=blue]{hyperref}
\usepackage{blkarray}
\usepackage{array}
\usepackage[dvipsnames]{xcolor}
\usepackage[normalem]{ulem}
\usepackage{wasysym}
\usepackage{multirow}
\usepackage{comment}
\usepackage{subcaption}
\usepackage{xspace}

\usepackage{tabularx}
\usepackage{array}   
\newcolumntype{L}{>{$}l<{$}} 
\newcolumntype{R}{>{$}r<{$}} 
\newcolumntype{C}{>{$}c<{$}} 

\usepackage{float}

\usepackage[capitalise]{cleveref}
\usepackage{placeins}

\date{\today}

\newcommand{\chemform}{\texorpdfstring{Sr\textsubscript{2}RuO\textsubscript{4}}{}}

\graphicspath{{./figures}}

\usepackage{comment}

\begin{document}
	\title{
        Why Scanning Tunneling Microscopy on \chemform{} sometimes doesn't  see the superconducting gap 
	}
	
	\author{Adrian Valadkhani}
    \email{valad@itp.uni-frankfurt.de}
	\affiliation{Institut f\"ur Theoretische Physik, Goethe-Universit\"at, 60438 Frankfurt am Main, Germany}
	\author{Jonas B.~Profe}
	\affiliation{Institut f\"ur Theoretische Physik, Goethe-Universit\"at, 60438 Frankfurt am Main, Germany}

	\author{Andreas Kreisel}
	\affiliation{Niels Bohr Institute, University of Copenhagen, DK-2200 Copenhagen, Denmark}
	
	\author{P.J. Hirschfeld}
	\affiliation{Department of Physics, University of Florida, Gainesville, Florida 32611, USA}
	
	\author{Roser Valent\'i}
    \email{valenti@itp.uni-frankfurt.de}
	\affiliation{Institut f\"ur Theoretische Physik, Goethe-Universit\"at, 60438 Frankfurt am Main, Germany}
	
	\date{\today}
	
	\begin{abstract}
        Scanning tunneling microscopy (STM) is perhaps the most promising way to detect the superconducting gap size and structure in the canonical unconventional superconductor Sr$_2$RuO$_4$ directly.
        However, in many cases, researchers have reported being unable to detect the gap at all in simple STM conductance measurements.
        Recently, an investigation of this issue on various local topographic structures on a Sr-terminated surface  found  that superconducting spectra appeared only in the region of small nanoscale canyons, corresponding to the removal of one RuO surface layer.
        Here, we analyze the electronic structure of various possible surface structures using first principles methods, and argue that bulk conditions favorable for superconductivity can be achieved when removal of the  RuO layer suppresses the RuO$_4$ octahedral rotation locally. We further propose alternative  terminations  to the most frequently reported Sr termination where superconductivity surfaces should be observed.
	\end{abstract}

	\maketitle
\section{Introduction}

The superconducting order parameter of  {Sr}$_2${RuO}$_4$~\citep{Mackenzie_RMP_2003,Maeno_JPSJ_2012,Kallin_IOP_2016,Mackenzie_npj_2017} is  thought to be of unconventional nature, but has proven unexpectedly difficult to identify.
Soon after its discovery in 1994, it was proposed as a promising candidate for a chiral $p$-wave, spin triplet  superconductor by analogy to superfluid $^3$He-A, based in particular on early evidence from NMR~\citep{Ishida_Nature_1998}.
Muon-spin rotation~\citep{Luke_Nature_1998} and Kerr effect~\citep{Xia_PRL_2006} measurements suggested intrinsic time-reversal symmetry (TRS) breaking~\citep{Rice_1995} below $T_c$, consistent with this proposal.
 Thermodynamic measurements provided clear evidence for low-energy quasiparticle states, however, 
suggesting the existence of gap nodes or deep minima~\cite{Bonalde_PRL_2000,Mao_JPSJ_2000,Maeno_PRL_2004,Hassinger_PRX_2017}.
\vskip .2cm

In 2019, the authors of Ref.~\cite{Pustogow_Nature_2019} challenged the chiral $p$-wave paradigm with in-plane $^{17}$O nuclear magnetic resonance measurements that found a significant  decrease in the {K}night shift below $T_\mathrm{c}$~\cite{Pustogow_Nature_2019,Ishida_correct}. Spin-triplet pairing was then definitively ruled out by comparison to the change of the entropy from earlier specific heat experiments~\citep{Brown_2020}. These measurements were accompanied by observations of shifts in the elastic constants~\cite{Benhabib2021,ghosh2020thermodynamic} together with experiments under strain~\cite{Grinenko2021} suggesting a two component nature of the order parameter.
All these results led to renewed theoretical attempts to  calculate   the superconducting ground state of   {Sr}$_2${RuO}$_4$  within a spin singlet picture~\citep{Astrid_PRL_2019,Scaffidi_PRR_2019,Gingras_PRL_2019,suh2019,Kaba19,Ramires2019,Acharya_CommPhys_2019,Kallin_PRB_2020, Romer2020, romer2020fluctuationdriven,Romer2021,kivelson_npj,clepkens2021,willa2021,Astrid3D2022,Merce2022,Henrik2022,palle2023constraints,hauck2023competition}, leading to a variety of proposals, including the even-parity 1D irreducible representation $B_{1g} (d_{x^2-y^2})$, multi-component orders such as 
$d_{x^2-y^2}+ig_{xy(x^2-y^2)}$ and $s'+id_{xy}$, as well as the 2D irreducible representation $E_{1g} (d_{xz}+id_{yz})$.   A two-component state is thought to be required to explain ultrasound measurements~\cite{Benhabib2021,ghosh2020thermodynamic} and recent $\mu$SR experiments under strain~\cite{Grinenko2021} (see however Ref.~\cite{Andersen2024,palle2023constraints}), but other recent measurements provide evidence for a single order parameter component~\cite{Li2021,Li2022}.

\vskip .2cm

While the experiments cited above and many others provide indirect evidence in support of one superconducting pairing channel or another, the community's ability to definitively identify the order parameter is severely hindered by the difficulty of making {\it direct} measurements of the superconducting gap over the Fermi surface.  The extremely small gap ($|\Delta|\leq350 \, \mu$eV)~\citep{Firmo2013,Madhavan_PNAS_2020} of the superconducting order parameter $\Delta(\mathbf{k})$ in {Sr}$_2${RuO}$_4$ means that angle-resolved photoemission (ARPES) experiments do not currently have the  fine energy and momentum resolution to detect spectral features reflecting the gap.  Scanning tunneling spectroscopy  experiments that detect  Bogoliubov quasiparticle interference (BQPI) provide, on the other hand, good momentum resolution and finer energy resolution in the best circumstances.  Recently, the BQPI technique was used for Sr$_2${RuO}$_4$~\citep{Madhavan_PNAS_2020}. This analysis, based on comparison with a simple lattice calculation of the joint density of states, suggested a $d_{x^2-y^2}$ superconducting gap symmetry for {Sr}$_2${RuO}$_4$.  A somewhat more sophisticated calculation involving first-principles surface Wannier functions was also compared to the same data~\cite{Bhattacharyya2023}, but more than one  gap function appeared to fit nearly equally well.   Nevertheless, BQPI appears to be the best possibility of ``directly" measuring the superconducting gap in this canonical unconventional superconductor.

\vskip .4cm

There is however one enduring, poorly understood puzzle regarding  scanning tunneling microscopy (STM) studies of superconducting Sr$_2$RuO$_4$.  
While some STM measurements have reported signatures of superconductivity on the surface of Sr$_2$RuO$_4$ for many years~\cite{Suderow2009,Firmo2013,Madhavan_PNAS_2020}, others on apparently equivalent surfaces under similar conditions are unable to detect any gap at all~\cite{kambara_scanning_2006, lupien_mk-stm_2011,Marques2021,Madhavan_private}.
Recognizing that this question was an important one to resolve in order to properly interpret STM data on the Sr$_2$RuO$_4$ surface,  the authors of Ref.~\cite{Olivares2022,Mueller2023} performed a systematic study of local STM conductance spectra at several distinct types of local topographic structures on the surface with different termination layers.
They found that superconducting spectra with coherence peaks appeared only in the region of small nanoscale ``pits" on the surface, corresponding to the removal of one RuO surface layer.
Since the Sr$_2$RuO$_4$ surface is thought to be reconstructed in a pattern of RuO$_6$ octahedral rotation similar to that of bulk Sr$_3$Ru$_2$O$_7$ and calcium doped bulk Sr$_2$RuO$_4$~\cite{Nakatsuji2000,Matzdorf2002,Marques2021,Kreisel2021}, it is natural to ask if the atomic layer on the surface might be electronically different to that placed immediately underneath it, such that conditions favorable to superconductivity are ``masked".
In other words, is it possible that superconductivity in Sr$_2$RuO$_4$ can never be observed on  a hypothetical perfect but reconstructed Sr$_2$RuO$_4$ surface, but is only revealed when such pits form?
Here we investigate the plausibility of such a scenario by performing first-principles-based electronic structure simulations.

\vskip .2cm

\vspace{0.2cm}
\section{Results}  
\vspace{0.2cm}

    \begin{figure}
		\centering
		\includegraphics[width=\linewidth,trim=0 0 0 0,clip]{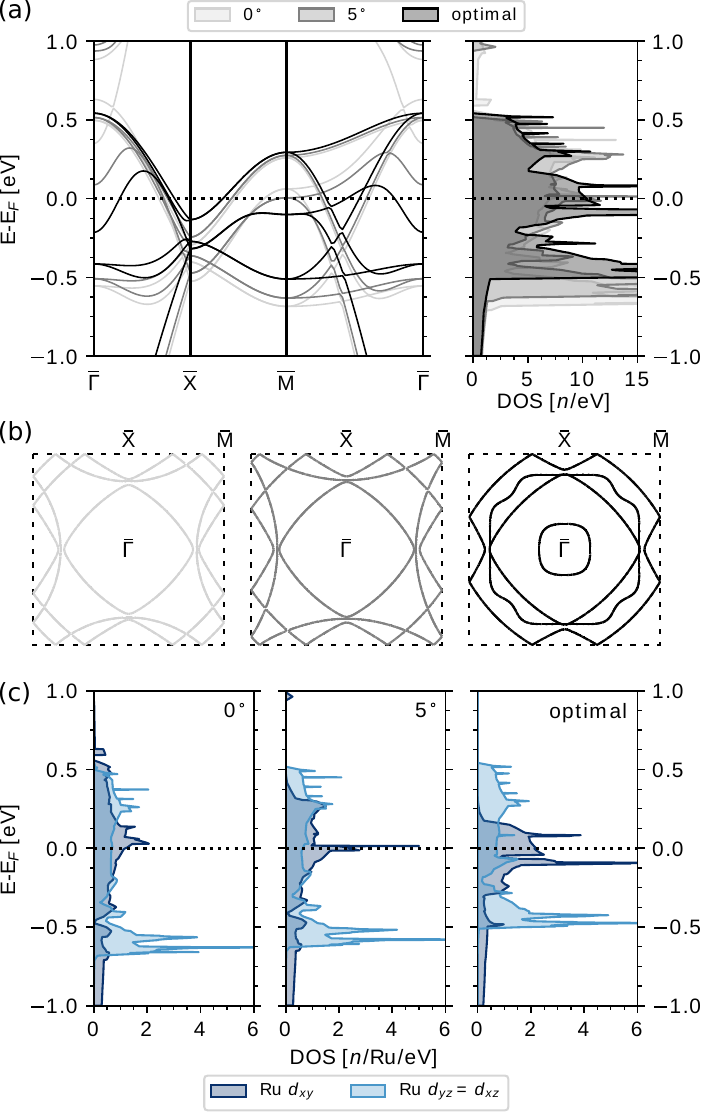}
		\caption{
			\textbf{Electronic structure of Sr Monolayer.}
            (a) Band structure along the high symmetry path of the folded Brillouin zone (BZ) in Sr$_2$RuO$_4$ and the density of states (DOS) for the rotations $0^{\circ}$ (lightgray), $5^{\circ}$ (gray) and the optimal one (black).
            (b) Fermi surfaces in the folded BZ 
            for the same three rotations. 
            We observe a Lifshits transition happening between the $5^{\circ}$ (gray) and the optimal rotation (black).
            (c) $d$ orbitally resolved DOS plotted  with different shades of blue.
            The $d_{xy}$ orbitals are the dominant orbitals at the Fermi level $E_F$.
        }
		\label{fig:cmp:rot:monolayer}
	\end{figure}
 
    In the following we examine how the surface termination and rotation angles of the RuO$_6$ octahedra in Sr$_2$RuO$_4$ affect its electronic structure employing density functional theory (DFT). 
    We  determine the optimal rotation angle depending on the type of termination, and perform large-scale canyon structure simulations to investigate the above scenario predicting that pits in the surface allow to observe signatures of bulk physics. 

    To understand better the changes induced by the octahedral rotation in \chemform{} we first focus on a \ce{Sr}-terminated monolayer for which we consider different RuO$_6$ rotation angles.
    The effect of the rotation on the band structure, Fermi surface, total density of states (DOS) and projected DOS is visualized in Fig.~\ref{fig:cmp:rot:monolayer} and is in agreement with previous reported results~\cite{chandrasekaran_engineering_2023}.
    As shown in Fig.~\ref{fig:cmp:rot:monolayer}, the octahedral rotation pushes the van-Hove singularity (vHs) -- located in the $d_{xy}$ orbital -- from a position above the Fermi level  at $0^{\circ}$ rotation, as in the bulk case~\cite{tamai_high-resolution_2019, Gingras_PRL_2019, romer2020fluctuationdriven}, to a position below the Fermi level at the optimal octahedral rotation value. 
    As a consequence of this, the $d_{xy}$-orbital dominated Fermi surface section near the $\bar{M}$-point  vanishes (see the band structure and Fermi surface plotted in Fig.~\ref{fig:cmp:rot:monolayer} (a)). 
    Furthermore, a pocket at the $\bar{\Gamma}$ point in the $d_{xy}$ orbital opens up at approximately $6^{\circ}$ octahedral rotation.
    This feature has been identified as a source of suppressed superconductivity in a recent theoretical investigation by some of the authors~\cite{Jonas2024}
    by performing functional renormalization group (FRG) calculations for wannierized tight-binding models for the surface of \chemform{}.
    
    In the same spirit and motivated by recent strain dependent experiments~\cite{Steppke_Science_2017, Watson_2018_strain, Grinenko2021, Jerzembeck2022, li2022elastocaloric} indicating a strong effect of the vHs position on $T_c$, we investigate different possible realizations of surfaces with the aim to recover bulk-like behavior in the surface layer, which would support the scenario that the detection of superconductivity on the surface is directly related to the surface having similar electronic properties as the bulk.
    By cleaving such surfaces, surface sensitive probes, like STS, as mentioned in the introduction, can then be utilized to image the superconducting order parameter. 
 
	\begin{figure}[!htp]
		\centering
		\includegraphics[width=1.0\linewidth]{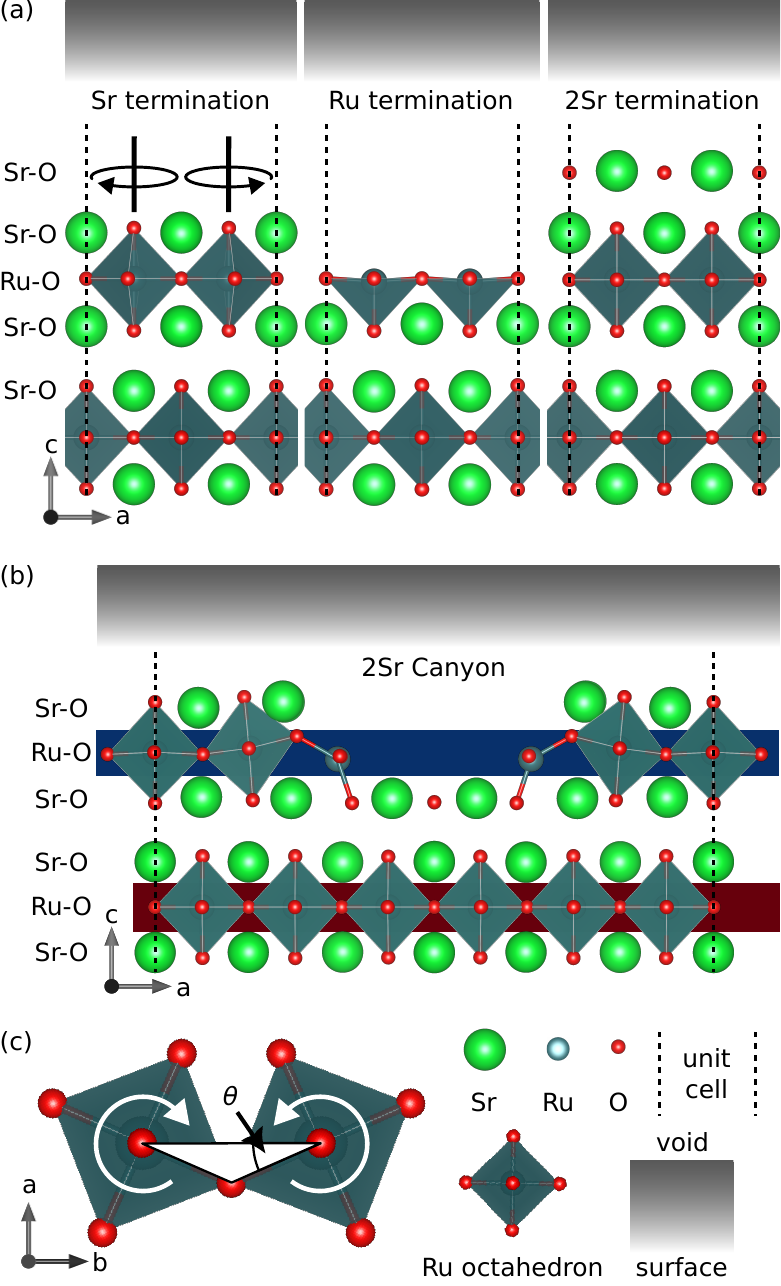}
        \caption{\textbf{Slab terminations} considered in this work. 
        (a-b) depict the surface and subsurface layers used for our slab calculations where the shaded gray region on top marks the position of the void for the calculations. The labels on the left  (Sr-O or Ru-O) denote the layer composition.
        Panel (a) shows the Sr-terminated, Ru-terminated and 2Sr-terminated slabs. 
        The rotation of RuO$_6$ octahedra is depicted on top of the Sr-termination figure.
        (b) Canyon-like structure with 2Sr termination in the pit (2Sr Canyon).
        This canyon-like structure has been reported in STM experiments~\cite{Olivares2022}.
        The dark blue and red horizontal bars are color indicators for Fig.~\ref{fig:canyon}.
        (c) Definition of the RuO$_6$ octahedra rotation angle.
        }
		\label{fig:terminations}
	\end{figure}
   
    We consider four different charge neutral surface configurations in Sr$_2$RuO$_4$ including the experimentally realized \ce{Sr} termination~\cite{Marques2021}, a \ce{Ru} termination, and a 2\ce{Sr} termination (see Fig.~\ref{fig:terminations} (a)). 
    While the latter two terminations, to our knowledge, have not yet been reported as a clean surface, our findings indicate that trying to cleave or grow these is highly desirable in order to restore bulk physics on the surface. 
    As a fourth configuration, we examine a canyon-like structure (Fig.~\ref{fig:terminations} (b)) similar to the suggested pit in~\cite{Olivares2022} and investigate how the electronic structure of \chemform{} is affected thereby. 
   
    \begin{figure}[b]
		\centering
		\includegraphics[width=1.0\linewidth]{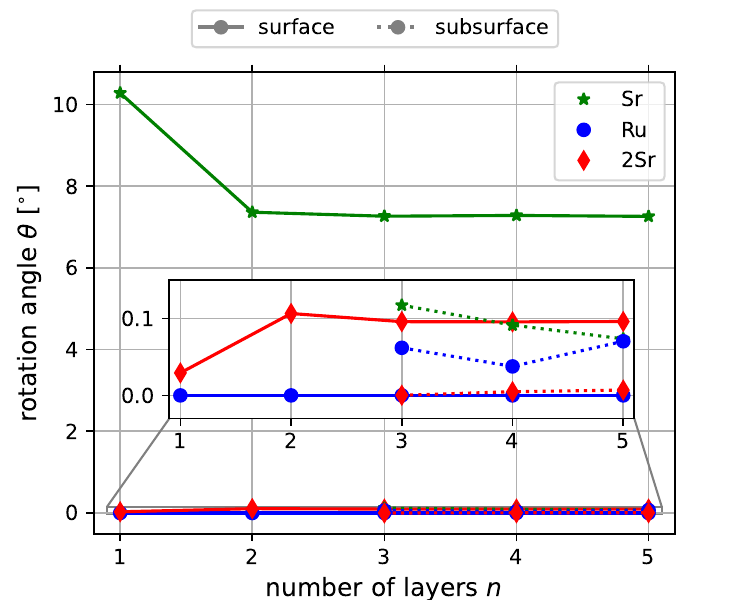} 
        \caption{
        \textbf{Octahedral rotation $\boldsymbol{\theta}$} (see Fig.~\ref{fig:terminations} (c) for the angle definition) as a function of 
        number of layers within the slab for both, surface (continuous line) and subsurface (dashed line) for all three terminations in
        Fig.~\ref{fig:terminations} (a). 
        Note that only as of a three-layer slab a subsurface layer exists.
        We find a much stronger surface reconstruction in the case of Sr termination (green solid curve) than for the Ru termination (blue, in inset) and the 2Sr termination (red, inset).
        All subsurface layers show negligible rotations.
        The exact values of the relaxation can be found in Tables \ref{tab:rotationAngle:SrT}, \ref{tab:rotationAngle:RuT} and \ref{tab:rotationAngle:2SrT}.
        }
		\label{fig:octrot}
	\end{figure}
 
    The optimized RuO$_6$ rotation angles for slabs with Sr surface termination (Fig.~\ref{fig:terminations} (a), left panel) are summarized in Fig.~\ref{fig:octrot} as a function of number of layers $n$ (see also Tab.~\ref{tab:rotationAngle:SrT}). 
    We find that the octahedral rotation in both surface and subsurface show rapid convergence to stable values  with the number of layers considered in the slab calculation, indicating the presence of small interlayer couplings,
    as expected from resistivity measurements~\cite{hussey_normal-state_1998}.
    Incorporating spin-orbit coupling (SOC) in the structural slab relaxations is not found to play a crucial role for the rotation angles, in contrast to the role it plays for the electronic properties near the Fermi surface~\cite{tamai_high-resolution_2019} where it was found to be essential. 
    In other words, even though to reproduce the experimental Fermi-surface shape spin-orbit coupling is essential, it has negligible effect on the structural relaxation which is the central focus of this study. 
    The optimal rotation angle of $7.17^{\circ}$ at the surface layer obtained in non-relativistic calculations is in agreement with experiments~\cite{Matzdorf2000,Matzdorf2002} and earlier theoretical investigations~\cite{Veenstra2013,chandrasekaran_engineering_2023}.
    Therefore, for the structure optimization we neglect the effect of SOC in multilayered slabs.
    Furthermore, we checked that structure optimization using different setups, like surface and subsurface optimization or structural optimization of the full slab with different depths, leads to changes in the rotation angle of the RuO$_6$ octahedra of at most $0.1^{\circ}$.
	

    We next consider slabs with Ru termination (Fig.~\ref{fig:terminations} (a) middle panel).
    In this case the surface RuO$_6$ octahedra  don't have top apex oxygens and the surface reconstruction after relaxation is less severe. The optimized rotation angles are therefore much smaller than in the previous case 
    as summarized in Fig.~\ref{fig:octrot} and in Tab.~\ref{tab:rotationAngle:RuT}. The slabs exhibit once again a fast convergence in rotation angles with number of layers considered within the slab (see inset of Fig.~\ref{fig:octrot}).

	\begin{figure*}[thb]
		\centering
		\includegraphics[width=\linewidth,trim=0 0 0 0,clip]{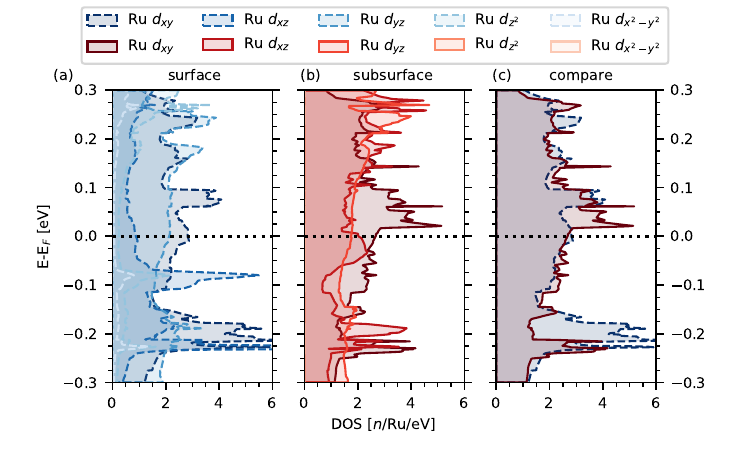}
		\caption{
			\textbf{DOS for the Ru atoms of the 2Sr canyon}
            plotted in three setups.
            The colors here correspond to the colors of the layers in Fig.~\ref{fig:terminations} (b) for the canyon.
            (a) shows the DOS of the top layer (blue, dashed) of the canyon for all $d$ orbitals.
            (b) It is the same as (a) but for the subsurface layer (red, solid).
            (c) Displays the comparison of the most dominant Ru orbital, $d_{xy}$, at the Fermi level for the surface and subsurface layer.
            The dos are normalized to one Ru atom.
        }
		\label{fig:canyon}
	\end{figure*}

 
    We also performed relaxations by considering a 2Sr termination (Fig.~\ref{fig:terminations} (a) right panel), where the surface is assumed to be terminated at a SrO layer, two layers away from the RuO plane. 
    In this case we observe, similarly to the Ru termination, a strong suppression of the RuO$_6$ octahedral rotation in the most outer RuO surface (see Fig.~\ref{fig:octrot} and Table~\ref{tab:rotationAngle:RuT}).
    The rotation angles again converge quickly with the number of layers within the slab.
    As expected from the electronic structure analysis of the effect of RuO$_6$ octahedral rotations in Sr$_2$RuO$_4$ (Fig.~\ref{fig:cmp:rot:monolayer}), we find bulk-like behavior in the absence of the octahedral rotation for both the 2Sr and Ru terminations.  

    We proceed now with a more complex slab termination.
    As reported in Ref.~\cite{Olivares2022} imperfections in the cleaving process of the crystal, which create small canyon-like structures, enable the observation of a superconducting state in STM.
    Here, we perform DFT calculations for one such case, a 2Sr canyon termination (Fig.~\ref{fig:terminations} (b)). 
    For that  we choose a trade-off between numerical feasibility and complexity of the structure.
    The canyon-like structure is built from previously relaxed Sr-terminated slabs.
    After removing the atoms in order to resemble the structures suggested in Ref.~\cite{Olivares2022}, we perform a relaxation of the whole structure.
    We find that near the edges of the canyon, a substantial displacement of the atoms occurs - while at the bottom of the pit, the RuO$_6$ octahedral rotation is nearly zero.
    Fig.~\ref{fig:canyon} shows the orbitally resolved DOS for the 2Sr canyon (Fig.~\ref{fig:terminations} (b)) where the contribution of the surface Ru is shown in shades of dashed blue (Fig.~\ref{fig:canyon} (a)) and the contribution coming from subsurface Ru is shown  in shades of solid red (Fig.~\ref{fig:canyon} (b)).
    Fig.~\ref{fig:canyon} (c) depicts the comparison of the $d_{xy}$ orbital projected DOS for the surface and subsurface.
    By this direct comparison it is visible that the subsurface layer has a clear bulk-like feature, with a clear peak close to the Fermi level (compare, for instance, with Ref.~\cite{Marques2021}, and with Fig.~\ref{fig:cmp:rot:monolayer} (c) left panel where the electronic structure of bulk Sr$_2$RuO$_4$ is emulated).
    As can be seen in Fig.~\ref{fig:canyon} the influence of the RuO$_6$ octahedral rotation is quite drastic - the electronic structure of the subsurface Ru layer retains the position of the bulk system's van-Hove singularity while the vHs of the surface Ru is shifted below the Fermi level, again resembling what we observed in the Sr-terminated slab (Fig.~\ref{fig:cmp:rot:monolayer}).
    This indicates that the subsurface electronic structure is the same as for bulk Sr$_2$RuO$_4$, since it is not affected by the octahedra rotation.
    Therefore, bulk-like behavior is expected, i.e.~the layer should show superconductivity when the tip is placed in the area of the canyon, as indicated by STM measurements~\cite{Olivares2022}. 
 
	\vspace{0.2cm}
	\section{ Discussion}
	\vspace{0.2cm}

    In this work, we investigated via DFT calculations the surface reconstruction of different charge-neutral surfaces of Sr$_2$RuO$_4$, including experimentally observed canyon structures.
    Our motivation was to understand the microscopic origin of contradictory reports of detection of superconductivity in STM measurements of Sr$_2$RuO$_4$.
    The results of our calculations are condensed into two main quantities which are (i) the structural properties in terms of the behavior of the RuO$_6$ octahedra at the surface and layers below and (ii) the (partial) density of states of the Ru-d states in the corresponding layers.
    To examine whether to expect a structure to have bulk-like behavior, we compare the  bulk and relaxed surface electronic structures. From experiments, we know that the bulk system is superconducting~\cite{Mackenzie_RMP_2003} while clean surface samples show no superconductivity~\cite{kambara_scanning_2006, lupien_mk-stm_2011,Marques2021,Madhavan_private}.
    In particular, we focus on the DOS of the $d_{xy}$ orbital since it is the main contributor to the DOS near the Fermi level, and to the van-Hove singularity driving the value of the $T_c$ as examined by strain experiments~\cite{Steppke_Science_2017, Watson_2018_strain, Grinenko2021, Jerzembeck2022, li2022elastocaloric}.
    It should be noted that at large octahedral rotation ($\theta$ $\geq$ 6$^{\circ}$), which is the case for experimentally reported Sr-terminated structures and found in our simulations, the van-Hove singularity lies below the Fermi level (E$_F$) and the DOS at E$_F$ is predominantly provided by the new $d_{xy}$ pocket opening near the $\Gamma$ point (see Fig.~\ref{fig:cmp:rot:monolayer} (b) right panel), in contrast to the bulk-like electronic structure (no RuO$_6$ rotation, Fig.~\ref{fig:cmp:rot:monolayer} (b) left panel) where this pocket is absent.
    Recent FRG calculations suggest that while carrying significant DOS, this new pocket is irrelevant for the pairing~\cite{Jonas2024}, and this might explain the absence of superconductivity at such surfaces.
    
 \vspace{0.2cm}   
    Using this qualitative understanding of how superconductivity is influenced by the structural state of the Sr$_2$RuO$_4$ and by examining the signatures of the density of states, we can indirectly connect the tendency towards superconducting pairing to the structural state of the  RuO$_6$ octahedra.
    We find that the density of states in the layers without rotation exhibits similar positions of the van Hove peaks as that of an unreconstructed (bulk) system; 
    in contrast to layers with significant octahedra rotation, thus connecting the electronic structures in the corresponding layers to the bulk electronic structure.
    Our calculations corroborate that for Sr-terminated surfaces, the RuO$_6$ octahedra show a rotation of  about $7^{\circ}$, in agreement with previous investigations~\cite{Matzdorf2000,Matzdorf2002,Veenstra2013,chandrasekaran_engineering_2023}.
    However, for both  2Sr- and Ru-terminated surfaces,  no significant octahedral rotation was found.

    This suggests that if these latter terminations could be successfully grown or cleaved, superconducting surfaces should be observed. 
    This in turn would allow for direct measurements of the superconducting order parameter utilizing Bogoliubov quasiparticle interference. 

    This suggestion is supported by our 2Sr canyon slab simulations resembling reported surface imperfections in Ref.~\cite{Olivares2022} where we found that the octahedral rotation in the subsurface is essentially absent.
    Accordingly, the electronic structure does not differ substantially from bulk - explaining why within these canyons a superconducting gap is observed in STM measurements~\cite{Olivares2022}.
    Therefore, we argue that STM (or other surface) experiments do not see the superconducting gap if they are located with their tip on the Sr-terminated surface with a reconstructed surface that contains octahedral rotation.

    While so far there has been no report of a clean 2Sr or Ru surface termination, our results suggest that fabricating such a termination could help significantly in determining the gap structure in \chemform{}. 
    Furthermore, this could settle the longstanding debate of whether there is a second order parameter or not~\cite{ghosh2020thermodynamic, Benhabib2021, Grinenko2021, li2022elastocaloric, palle2023constraints, mueller2023constraints, Li2021}. 
    While a relatively clean surface might be required to obtain Bogoliubov quasiparticle interference patterns, the observation of a STM gap in canyon defects already allows for a more thorough investigation of the gap structure.
	
	\section{Methods}
	
	\vspace{0.2cm}{\bf First principles calculations}

    \vspace{0.2cm}
    
    We performed ab initio electronic structure calculations within density functional theory (DFT) ~\cite{Hohenberg1964,Kohn1965} by using the Vienna Ab initio Simulation Package (VASP)~\cite{Kresse1993,Kresse1996,Kresse1996_2} within the pseudo-potential augmented plane-wave\cite{Bloechl1994,Kresse1999} (PAW) basis set.
    The calculations were performed with the Perdew–Burke–Ernzerhof (PBE)~\cite{Perdew1996} exchange correlation functional as a generalized gradient approximation (GGA), and a plane-wave cutoff of $800$\,eV was chosen. 
    First the bulk structure was relaxed on a $ 16\times16\times4$ $k$-point mesh using the conventional unit cell of \chemform{}.
    The relaxed unit cell was then transformed to a tetragonal $\sqrt{2}\times\sqrt{2}\times1$ unit cell in order to have two inequivalent Ru sites and four inequivalent O sites in the $ab$ plane. 
    The convergence criterion was chosen to be $2\cdot 10^{-3} eV/$\AA{} for these slabs and $1\cdot 10^{-3} eV/$\AA{} for the bulk.
    
    From the optimized bulk unit cell we generated slabs with up to 5 Ru layers with Sr, 2Sr or Ru termination on both ends.
    A void of $15$\,\AA{} was set on top of the surfaces.
    In these slab unit cells we rotated the RuO$_6$ octahedra starting form their initial position, i. e. $180^{\circ}$ between two neighboring Ru atoms and the O in between, by adding a displacement $\textbf{d}_{\text{O}}$ for the O atoms in the Ru plane dependent on the rotation angle $\theta$
    \begin{equation}
        \textbf{d}_{\text{O}}(\theta)=
        \begin{pmatrix}
            0                & \tan(\theta) & 0 \\
            -\tan(\theta)    & 0            & 0 \\
            0                & 0            & 0
        \end{pmatrix}
        \cdot \textbf{r}_{\text{O}},
    \end{equation}
    where $\textbf{r}_{\text{O}}$ is the initial positions of the O atoms in the Ru plane.
    For all slabs we optimized every pair of Ru layers symmetrically.
    Crosscheck for non symmetric setups were done, however not found to be optimal.
    
    The scheme of optimizing was done carefully for all terminations as follows:
    (i) We first started with one Ru layer.
    (ii) This was optimized first with a low resolution energy landscape.
    (iii) Then we relocated and zoomed in to restart with a higher resolution.
    (iv) We repeated this procedure until we had a sufficient amount of resolution.
    (v) As a last step of refinement we let VASP internally relax, to printout the forces and confirm that we are in a minimum.
    (vi) Restart with an additional Ru layer, and use the previous optimal rotation angle as new starting rotation angle.
    The $k$ mesh for optimizing these structures was $6\times6\times1$.\\
    For the 2Sr canyon structure, we started from the relaxed 3 layer Sr-terminated slab and expanded the unit cell by $\sqrt{2}\times\sqrt{2}\times1$. 
    This supercell was then extended in one direction $3\times1\times1$.
    Finally we dug a hole on both surfaces and started a relaxation.
    The forces were reduced to $5\cdot10^{-3} eV/$\AA{} with a $k$-mesh of $2\times6\times1$.

	\vspace{0.2cm}
	
	\section{Acknowledgments}

    The authors are grateful to S.~Mukherjee for valuable discussions in the early stages of this project and V.~Madhavan for providing valuable information concerning STM measurements of Sr$_2$RuO$_4$.
	R.V., J.B.P.~and A.V.~gratefully acknowledge support by the Deutsche Forschungsgemeinschaft (DFG, German Research Foundation) for funding through 
	Project No. 411289067 (VA117/15-1) and TRR 288 — 422213477 (project A05). 
    A.K.~acknowledges support by the Danish National Committee for Research Infrastructure (NUFI) through the ESS-Lighthouse Q-MAT.  P.J.H.~acknowledges support from NSF-DMR-2231821.

	\subsubsection{Author Contributions}
    All authors contributed to the data analysis, discussions and writing of the manuscript. A.V. performed the DFT simulations.
 
	\subsubsection{Data availability}
    All simulation data is available upon reasonable request.
    
\bibliography{Sr2RuO4}
	
	\clearpage
	
	\widetext
	\appendix
	\begin{center}
		\textbf{\large \textit{Supplementary Information}:\\ \smallskip Why STM on \chemform{} (sometimes) doesn't  see the gap}
        \bigskip
	\end{center}
	\twocolumngrid
	
	\setcounter{equation}{0}
	\setcounter{figure}{0}
	\setcounter{table}{0}
	\setcounter{page}{1}
	
	\makeatletter
	\renewcommand{\theequation}{S\arabic{equation}}
	\setcounter{figure}{0}  
    \renewcommand{\thefigure}{S\arabic{figure}}
	\renewcommand{\thetable}{S\Roman{table}}
	
	\DeclareFontShape{OT1}{cmss}{m}{it}{<->ssub*cmss/m/sl}{}
	\renewcommand{\rmdefault}{cmss}
	\renewcommand{\sfdefault}{cmss}
	
	\DeclareFontFamily{OT1}{cmbr}{\hyphenchar\font45 }
	\DeclareFontShape{OT1}{cmbr}{m}{n}{%
		<-9>cmbr8
		<9-10>cmbr9
		<10-17>cmbr10
		<17->cmbr17
	}{}
	\DeclareFontShape{OT1}{cmbr}{m}{sl}{%
		<-9>cmbrsl8
		<9-10>cmbrsl9
		<10-17>cmbrsl10
		<17->cmbrsl17
	}{}
	\DeclareFontShape{OT1}{cmbr}{m}{it}{%
		<->ssub*cmbr/m/sl
	}{}
	\DeclareFontShape{OT1}{cmbr}{b}{n}{%
		<->ssub*cmbr/bx/n
	}{}
	\DeclareFontShape{OT1}{cmbr}{bx}{n}{%
		<->cmbrbx10
	}{}

	\titleformat{name=\section}
	{\normalfont\large\bfseries\MakeUppercase}{\MakeUppercase{\thesection}}{0pt}{}
	\titleformat{name=\subsection}
	{\normalfont\bfseries}{\thesection}{0pt}{}
	\titlespacing{\section}{0cm}{0.7cm}{0.01cm}
	\titlespacing{\subsection}{0cm}{0.45cm}{0cm}
	
	\title{
		Supplementary Information: \\ Why STM on Sr$_2$RuO$_4$ (sometimes) doesn't  see the gap
	}
	\author{Adrian Valadkhani}
	\affiliation{Institut f\"ur Theoretische Physik, Goethe-Universit\"at, 60438 Frankfurt am Main, Germany}
	\author{Jonas Profe}
	\affiliation{Institut f\"ur Theoretische Physik, Goethe-Universit\"at, 60438 Frankfurt am Main, Germany}

	\author{Andreas Kreisel}
	\affiliation{Niels Bohr Institute, University of Copenhagen, DK-2200 Copenhagen, Denmark}
	
	\author{Peter Hirschfeld}
	\affiliation{Department of Physics, University of Florida, Gainesville, Florida 32611, USA}
	
	\author{Roser Valent\'i}
	\affiliation{Institut f\"ur Theoretische Physik, Goethe-Universit\"at, 60438 Frankfurt am Main, Germany}
	
	\subsection*{Supplementary Note 1: Tables}
    \begin{table}[ht]
    \begin{tabular}{|c||c||l|l|l|l|l|}
    \hline
    No. Ru & type    & \multicolumn{1}{c|}{RuL1} & \multicolumn{1}{c|}{RuL2} & \multicolumn{1}{c|}{RuL3} & \multicolumn{1}{c|}{RuL4} & \multicolumn{1}{c|}{RuL5} \\ \hline\hline
    1      & all & 10.2837 &        &        &        &        \\ \hline \hline
    2      & all & 7.3645  & 7.3644 &        &        &        \\ \hline \hline
    3      & sur & 7.2825  & 0      & 7.2845 &        &        \\ \hline
    3      & all & 7.2672  & 0.1171 & 7.2662 &        &        \\ \hline \hline
    4      & sur & 7.1673  & 0      & 0      & 7.1721 &        \\ \hline
    4      & all & 7.2929  & 0.0916 & 0.091  & 7.2789 &        \\ \hline \hline
    5      & sur & 7.1709  & 0      & 0      & 0      & 7.1719 \\ \hline
    5      & sub & 7.2518  & 0.0959 & 0      & 0.0961 & 7.2519 \\ \hline
    5      & all & 7.2626  & 0.0734 & 0.0742 & 0.0735 & 7.2632 \\ \hline
    \end{tabular}
    \caption{
    Here we list the rotation angles in ${}^{\circ}$ in dependence of the number of Ru layers (No. Ru) for 1 layer (RuL1) to 5 layers (RuL5) with Sr termination. 
    Each layer has several rows, which correspond to either a full relaxation (all), only surface (sur), surface and subsurface (sub).
    }
    \label{tab:rotationAngle:SrT}
\end{table}

\begin{table}[h]
    \begin{tabular}{|c||c||l|l|l|l|l|}
    \hline
    No. Ru & type    & \multicolumn{1}{c|}{RuL1} & \multicolumn{1}{c|}{RuL2} & \multicolumn{1}{c|}{RuL3} & \multicolumn{1}{c|}{RuL4} & \multicolumn{1}{c|}{RuL5} \\ \hline\hline
    1      & all & 0  &         &        &        &   \\ \hline \hline
    2      & all & 0  & 0       &        &        &   \\ \hline \hline
    3      & sur & 0  & 0       & 0      &        &   \\ \hline
    3      & all & 0  & 0.0618  & 0      &        &   \\ \hline \hline
    4      & sur & 0  & 0       & 0      & 0      &   \\ \hline
    4      & all & 0  & 0.0378  & 0.0375 & 0      &   \\ \hline \hline
    5      & sur & 0  & 0       & 0      & 0      & 0 \\ \hline
    5      & sub & 0  & 0.0699  & 0      & 0.0696 & 0 \\ \hline
    5      & all & 0  & 0.0702  & 0.0001 & 0.0704 & 0 \\ \hline
    \end{tabular}
    \caption{
    Here we list the rotation angles in ${}^{\circ}$ in dependence of the number of Ru layers (No. Ru) for 1 layer (RuL1) to 5 layers (RuL5) with Ru termination. 
    Each layer has several rows, which correspond to either a full relaxation (all), only surface (sur), surface and subsurface (sub).
    }
    \label{tab:rotationAngle:RuT}
\end{table}

\begin{table}[h]
    \begin{tabular}{|c||c||l|l|l|l|l|}
    \hline
    No. Ru & type    & \multicolumn{1}{c|}{RuL1} & \multicolumn{1}{c|}{RuL2} & \multicolumn{1}{c|}{RuL3} & \multicolumn{1}{c|}{RuL4} & \multicolumn{1}{c|}{RuL5} \\ \hline\hline
    1      & all & 0.0292 &        &        &        &        \\ \hline \hline
    2      & all & 0.106  & 0.1067 &        &        &        \\ \hline \hline
    3      & sur & 0.0883 & 0      & 0.0881 &        &        \\ \hline
    3      & all & 0.0957 & 0.0002 & 0.0958 &        &        \\ \hline \hline
    4      & sur & 0.0993 & 0      & 0      & 0.0993 &        \\ \hline
    4      & all & 0.0954 & 0.0045 & 0.0051 & 0.0956 &        \\ \hline \hline
    5      & sur & 0.0882 & 0      & 0      & 0      & 0.0881 \\ \hline
    5      & sub & 0.0964 & 0.0009 & 0      & 0.0007 & 0.0961 \\ \hline
    5      & all & 0.0959 & 0.0069 & 0.0001 & 0.0069 & 0.096  \\ \hline
    \end{tabular}
    \caption{
    Here we list the rotation angles in ${}^{\circ}$ in dependence of the number of Ru layers (No. Ru) for 1 layer (RuL1) to 5 layers (RuL5) with 2Sr termination. 
    Each layer has several rows, which correspond to either a full relaxation (all), only surface (sur), surface and subsurface (sub).
    }
    \label{tab:rotationAngle:2SrT}
\end{table}
	\begin{figure}[!ht]
		\centering
		\includegraphics[width=\linewidth,trim=0 0 0 0,clip]{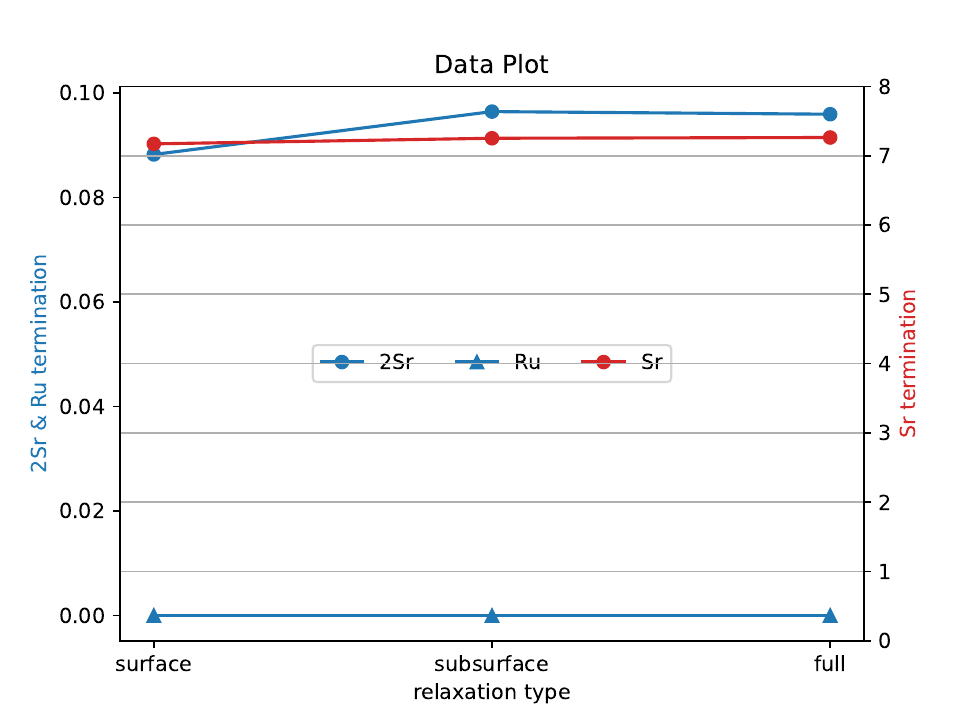}
		\caption{
			\textbf{Rotation angle} in dependence of freedom during the relaxation in DFT for all 3 terminations of the 5 layer slabs. 
        }
		\label{fig:5layer:relaxtype:convergence}
	\end{figure}
	
    Here we list all tables containing the optimized rotation angles from our DFT calculations with VASP.
    The Sr termination is in \ref{tab:rotationAngle:SrT}, the Ru termination in \ref{tab:rotationAngle:RuT} and the 2Sr termination in \ref{tab:rotationAngle:2SrT}.
    \cref{fig:5layer:relaxtype:convergence} shows the dependence of the rotation angle on the freedom during the relaxation.
    This plot is generated for the 5 layer case and benchmarks the rotation angle of the RuO$_6$ octahedra for the surface, by letting the structure relax only at the surface (labeled as: surface), using subsurface and surface (labeled as: subsurface) and also by letting the full structure relax (labeled as: full).
    The rotation angle converges very fast, showing very little deviation between surface and the other relaxation types.

	\vspace{0.2cm}
	\vspace{0.2cm}{\bf Comparison of bulk, monolayer, and canyon}
	\vspace{0.2cm}
 
	\begin{figure}[]
		\centering
		\includegraphics[width=\linewidth]{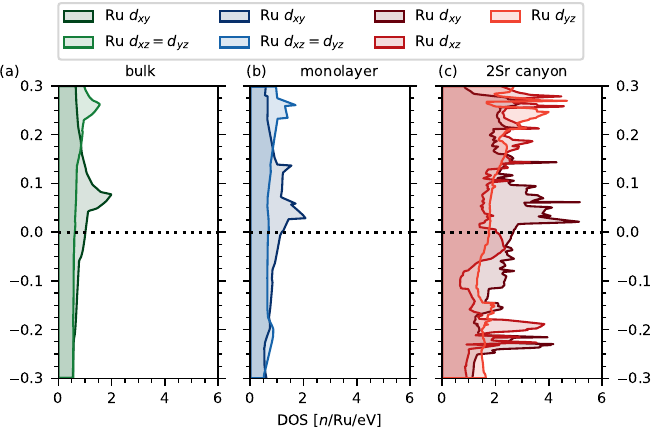}
		\caption{
			\textbf{Comparison of the DOS}
            of bulk, Sr-termination monolayer and 2Sr-canyon for the Ru $d_{xy}$ orbital to verify, that all of them have the same bulk like feature.
        }
		\label{fig:compare:bulk:monolayer:canyon}
	\end{figure}
 
    Here we show that the subsurface of the 2Sr canyon DOS has bulk-like features.
    In \cref{fig:compare:bulk:monolayer:canyon} we visualize the DOS of the $d_{xy}$ Ru orbital.
    For all three cases it shows a significant peak between 0.025 to 0.1 eV, which corresponds to the vHs in the $d_{xy}$ orbital.

	\vspace{0.2cm}
	\vspace{0.2cm}{\bf Ru canyon electronic structure}
	\vspace{0.2cm}

    \begin{figure}[]
    \centering
    \begin{subfigure}[]{\linewidth}
        \caption{}
        \includegraphics[width=\linewidth,clip=True,trim=0 0 0 35]{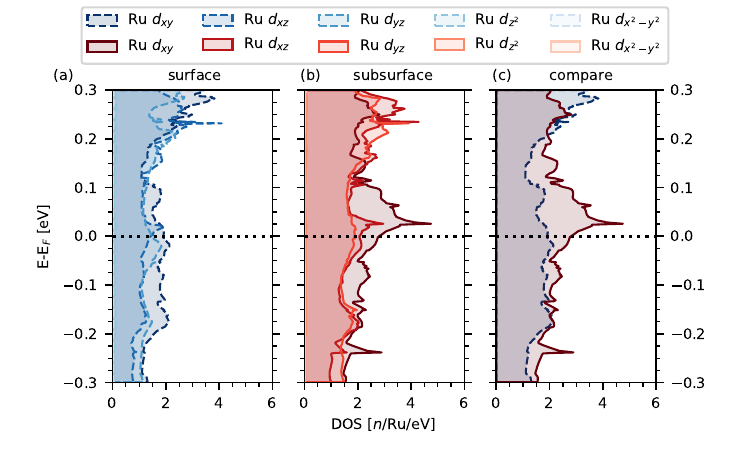}
    \end{subfigure}
    
    \begin{subfigure}[]{\linewidth}
        \caption{}
        \includegraphics[width=\linewidth,clip=True,trim=0 800 0 0]{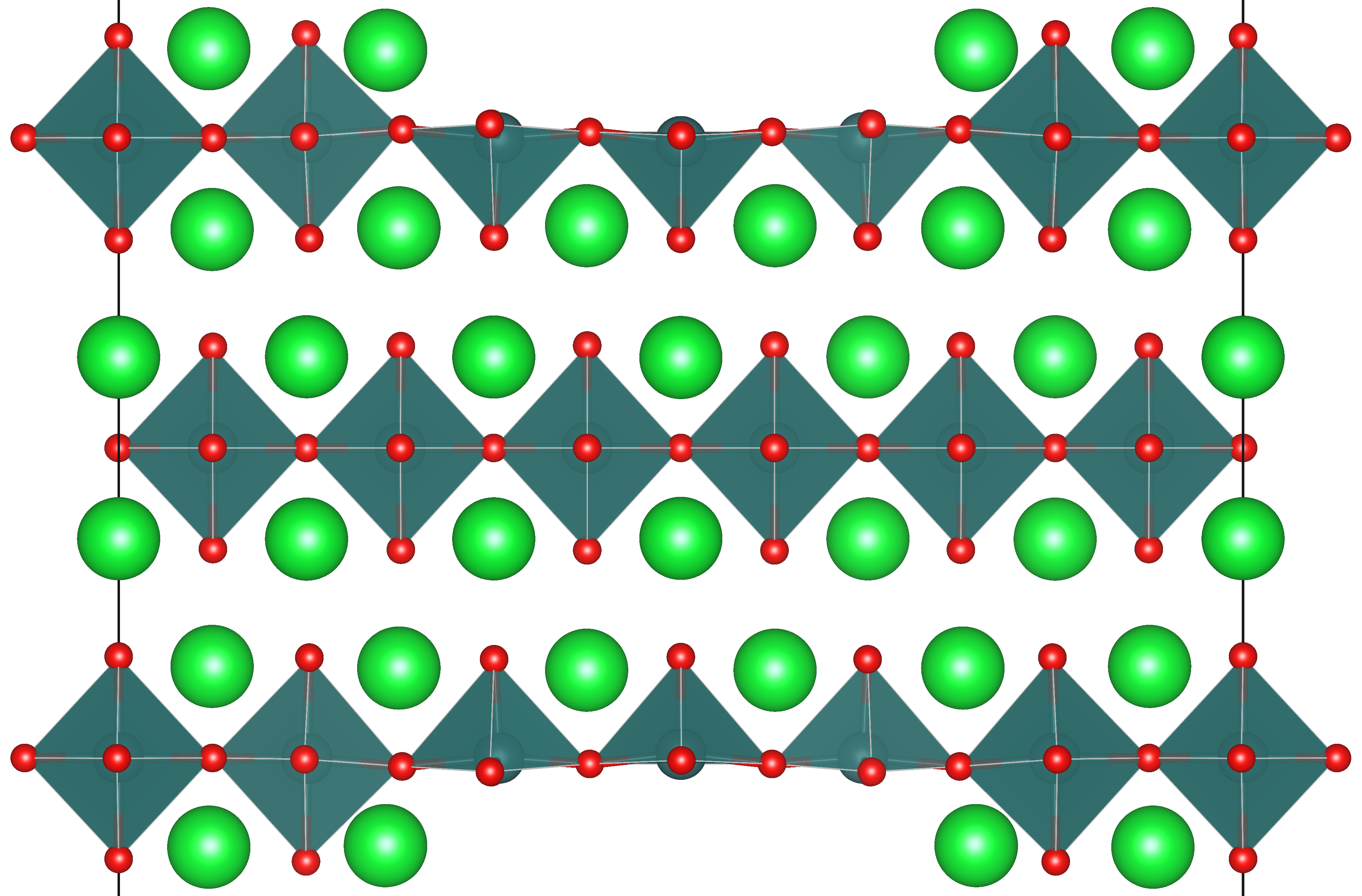}
    \end{subfigure}
    \caption{\textbf{Results for Ru canyon} 
    (a) Comparison between surface and subsurface, 
    and (b) the Ru canyon.}
    \label{fig:RuCanyon}
    \end{figure}
    A second pit proposed by the experiments in \cite{Olivares2022} is depicted in \cref{fig:RuCanyon}.
    Instead of a 2Sr termination in the pit one has a Ru termination.
    Similar to the 2Sr canyon (\cref{fig:canyon} main text) we decomposed the DOS in surface and subsurface layer.
    The DOS of the surface with the Ru octahedra without the top apex oxygen is very low.
    One the other hand one can see a nice agreement of the DOS for the subsurface layer and the bulk, which has much higher DOS values.
    Only the subsurface layer shows the expected bulk-like DOS.
    Unfortunately, the DOS of the subsurface might be screened by the Ru layer in the pit, therefore, visualizing the contributions of this canyon experimentally might be more challenging.
    
\end{document}